\documentclass[a4paper]{article}
\usepackage[T1]{fontenc}
\usepackage[utf8]{inputenc}
\DeclareUnicodeCharacter{0301}{\'{e}}

\usepackage{graphicx}
\usepackage{hyperref}
\hypersetup{
  setpagesize  = false,
  colorlinks   = true,    
  urlcolor     = blue,    
  linkcolor    = black,   
  citecolor    = black    
}
\usepackage{authblk}
\usepackage{geometry}
\usepackage{setspace}
\usepackage{verbatim}
\usepackage{amsmath}
\usepackage{amssymb}

\usepackage{verbatim} 
\usepackage{ragged2e} 

\usepackage{amsfonts}
\usepackage{cancel}
\usepackage{booktabs}

\newcommand{\keywordsenglishname}{Keywords}

\renewenvironment{abstract}{%
        \begin{center}
	\begin{minipage}{14cm}
	{\textbf{\abstractname:}}
}{
        \end{minipage}
	\end{center}
}
\newenvironment{abstractinenglish}{
        \def\abstractname{\abstractinenglishname}
	\begin{abstract}
}{
        \end{abstract}
}

\newenvironment{keywordsenglish}{
        \def\abstractname{\emph{\keywordsenglishname}}
	\begin{abstract}
}{
        \end{abstract}
}

\title {
Magnetic superconductivity
\\[1ex] 
\large 
}\author{David Möckli \href{https://orcid.org/0000-0002-4528-9755}{\includegraphics[scale=0.04]{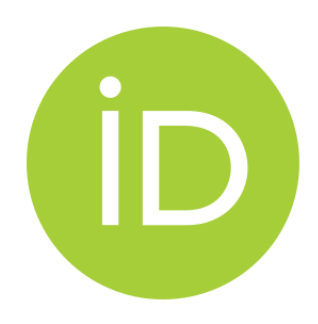}} }
\author{Murilo Kessler de Azambuja \href{https://orcid.org/0009-0002-5545-327X}{\includegraphics[scale=0.04]{orcidicon.png}} }

\affil{
Universidade Federal do Rio Grande do Sul, Instituto de Física,  91501-970, Porto Alegre,  RS, Brasil.}
\date{}

\usepackage{fancyhdr}
\begin{document}

\maketitle
\vspace{6pt}

\begin{abstractinenglish}
\emph{
This paper serves as a primer on superconductivity, inviting students for further investigation. Although the theory of superconductivity is a many-body quantum theory, here we take a more didactic route based on thermodynamics and symmetry. We briefly survey the more than a century-old field and provide a one-sentence definition of a superconductor. Surprisingly, many textbooks lack such a definition, usually introducing superconductors through their properties rather than by definition. We explain the concept of an order parameter, symmetry, and symmetry breaking. Based on this, we clarify the difference between conventional and unconventional superconductors, which is frequently a confusing topic for newcomers. We provide the reader with a taste of a current research topic in the field of unconventional superconductivity and magnetism. For this, we explain the concept of time-reversal symmetry breaking in condensed matter physics, which is usually associated with a form of magnetism. Here, we show that time-reversal symmetry might be broken in superconductors, leading to magnetic properties due to superconductivity itself. For this purpose, we utilize the method of the Ginzburg-Landau theory for phase transitions. We discuss the field of chiral superconductivity and guide the interested reader to further study.
}
\end{abstractinenglish}

\begin{keywordsenglish}
\emph{superconductivity, symmetry, time-reversal} 
\end{keywordsenglish}

\section{Introduction}

This paper is based on an invited talk, originally titled \textit{Magnetic Superconductivity}, presented by one of the authors at the XI Brazilian School of Magnetism, held in November 2023 in Porto Alegre, Brazil. The primary audience comprised undergraduate and graduate students without previous knowledge of superconductivity. 
Both the talk and paper target students aspiring to delve into the field of unconventional superconductivity, providing them with initial exposure to the subject and guiding them towards relevant references in the field.

The prerequisite to follow this paper is a basic understanding of thermodynamic potentials, such as free energy, and introductory knowledge of the Schrödinger equation. 
Superconductivity is inherently a quantum mechanical phenomenon that lacks a classical equivalent.  
However, by accepting a few definitions that we introduce here, we suspect that it is already possible to get a feel of how it is to develop a simple project in the field of theory of superconductivity. 
This approach allows for a smoother transition into the more technical aspects of superconductivity, making the subject more accessible to those with a preliminary background in related fields.

Textbooks often introduce superconductors through some of their non-universal properties, frequently leaving clear definitions wanting. This issue becomes more pronounced in the burgeoning field of unconventional superconductivity, where introductory texts can appear particularly arcane. To address this, we seize the opportunity to provide clear definitions of conventional and unconventional superconductivity. While a thorough understanding of these topics typically necessitates a grounding in group theory and the microscopic Bardeen-Cooper-Schrieffer (BCS) theory, we choose to bypass these prerequisites, adopting a colloquium-style approach. 
This allows students to assess whether pursuing a more advanced project in superconductivity aligns with their interests, providing a more accessible entry point into the subject without sacrificing depth.

Albeit its cutting-edge appeal, superconductivity was discovered more than a century ago in 1911, a decade before the formulation of the Schrödinger equation. 
The first successful quantum theory of superconductivity called the Bardeen-Cooper-Schrieffer (BCS) theory, had to await almost half a century since the discovery, which was published in 1957 \cite{bcs1957}. 
Now, almost 70 years after BCS theory, there is still no consensus on the cause (or causes) of superconductivity in many materials. 
However, the situation differs from 1950, when theoretical proposals were lacking. Today, there is a menu of apparently good theories that await critical experimental tests. 

Although BCS theory is a microscopic quantum theory, it abstains from specifying the cause of superconductivity, which we call the \textit{pairing mechanism} in the technical literature. BCS works for almost all mechanisms, which explains its success and universality. Two partial reasons for the slow progress in mechanism research are that mechanism theory has a reputation of being hard, and designing experiments perhaps even harder. 
Because of the difficulty of studying the root cause of superconductivity, two schools of study emerged in the field, which has been reasonably well-identified since 1990. 
The first school are the \textit{mechanism people}, which apply sophisticated mathematical techniques borrowed from high-energy physics to elucidate the mechanisms of superconductivity. Nowadays, we call these techniques \textit{condensed matter field theory}. 
This important field has a steep learning curve, and despite its remarkable progress in recent years, the detailed technical understanding still eludes many in the field.
The second school are the \textit{symmetry people}. Instead of focusing on specific causes, this school recognized that one may narrow down the menu of possibilities by studying the symmetry properties of the materials. This allows one to formulate impossibility theorems, which is a powerful and easy-to-learn tool. 
In this paper, we aim to introduce the reader to the second school. While we will touch upon pairing mechanisms, they will be presented as given.

In Sec. \ref{sec:what}, we explore where one finds superconductors and the significance of their study. 
Following this, Sec. \ref{sec:order_parameter} lays the foundation for superconductivity, axiomatically introducing the necessary quantum mechanical concepts. 
In Sec. \ref{sec:unconventional}, we delve into the crucial symmetries necessary to understand both conventional and unconventional superconducting states. Sec. \ref{sec:gl} employs a comparison with ferromagnetism to elucidate the Landau theory of spontaneous symmetry breaking. 
Finally, in Sec. \ref{sec:two}, we construct the basic theory of a two-component superconductor, revealing three thermodynamically distinct superconducting phases, including a chiral phase characterized by its exotic magnetic properties. 
We conclude with a brief discussion in Sec.  \ref{sec:discussion}, and provide a guide of references for the interested.

\section{Where, when and what? \label{sec:what}}

\subsection{Where?}

Many of us are familiar with the visually stunning film Avatar from 2009, featuring the breathtaking landscape of the Hallelujah Mountains, inspired by China's Zhangjiajie National Forest Park. In the movie, these towering mesa-like formations are composed of a fictional element called \textit{unobtanium}, which serves as an ambient pressure and temperature superconductor on the planet Pandora. These superconducting islands possess the remarkable ability to memorize the magnetic flux that threads through them, allowing them to levitate at a fixed height above the ground. While this scene captivates audiences with its science fiction allure, the underlying physical phenomenon is anything but fictional. In reality, however, humanity has not yet discovered how to engineer ambient temperature and pressure superconductors. Presently, our superconducting materials require frigid temperatures, typically below $-200^\circ$C; see Fig. \ref{fig:history_of_sc} for a sample of superconducting materials.

Unfortunately, superconductors on Earth are not as plentiful as those on Pandora, but they certainly exist. Where can we find them? As of the writing of this paper, superconductivity finds its most prominent applications in niche fields requiring the generation of powerful magnetic fields. These applications include magnetic resonance imaging (MRI) technology, ubiquitous in modern hospitals, magnetic levitation systems utilized in high-speed trains, and particle accelerators.
The astute reader might wonder: what about energy transmission? Aren't superconductors incredibly efficient electrical conductors? Indeed. Not long ago, many researchers would motivate their research proposals envisioning a future where superconductors would revolutionize the power grid. So why, over a century after the discovery of the first superconductor, are our power cables still not superconducting? As often occurs in fundamental research, the practical outcomes don't always align with initial expectations, but in the long run, often surpasses them. Currently, due to the significant cooling requirements, superconducting energy transmission isn't the most economically viable option for widespread implementation. Consequently, superconductors remain primarily employed in specialized applications. 

\subsection{When?}

\begin{figure}
    \centering
    \includegraphics[width = \textwidth]{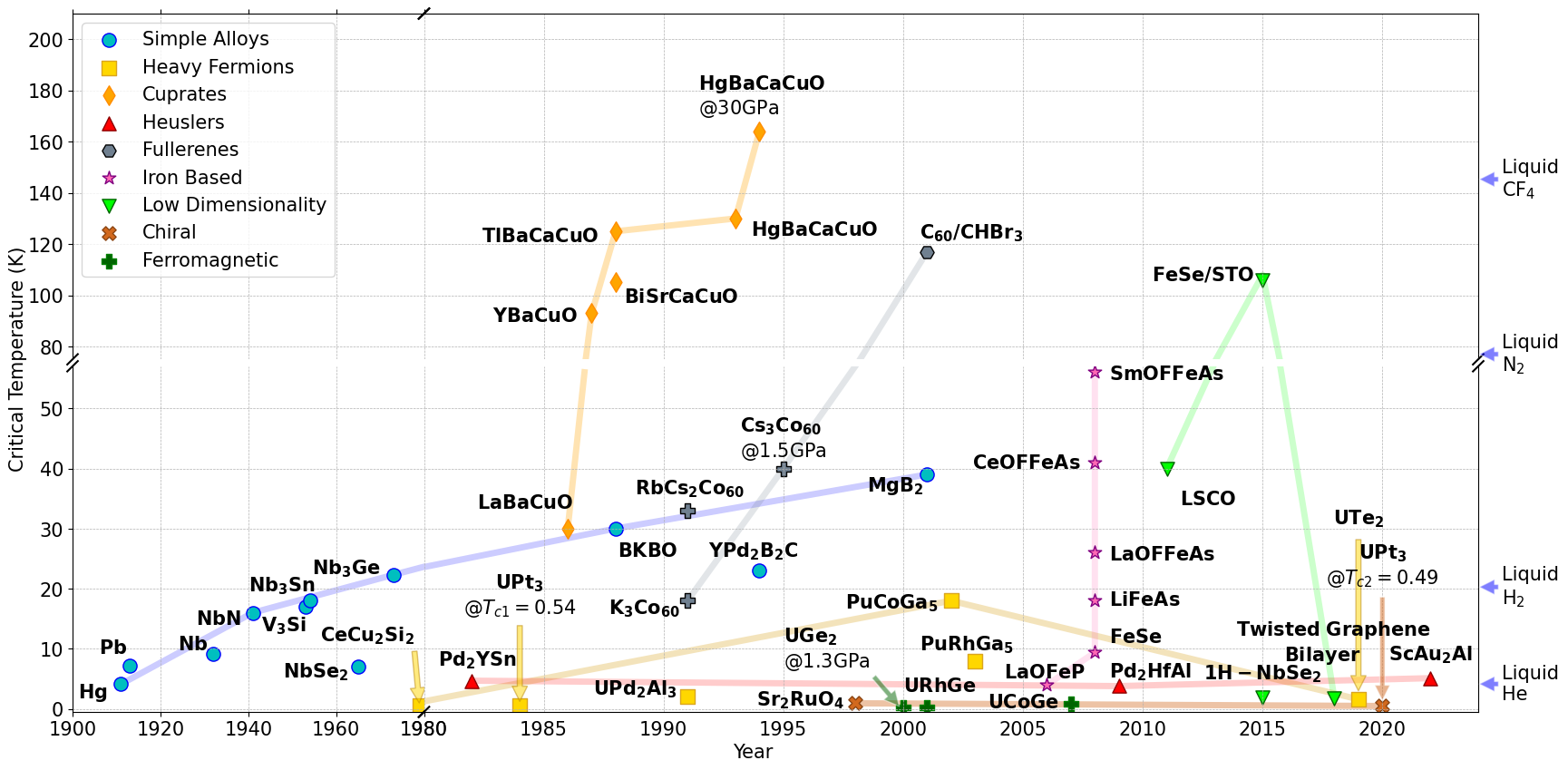}
    \caption{
    Historical overview of the discovery dates of superconducting materials and their respective critical temperatures, separated by "families." 
    }
    \label{fig:history_of_sc}
\end{figure}

In Fig. \ref{fig:history_of_sc}, we present a sample of superconductors categorized by their year of discovery and critical temperature, with each symbol in the legend representing a different family. The blue circles $(\circ)$ denote simple metallic alloys, which constituted the first family discovered and were predominantly explained by electron-phonon interaction. These alloys held sway in the field for approximately half a century. However, in 1986, ceramic copper and oxygen-based superconductors, known as \textit{cuprates}, were serendipitously discovered. Cuprates currently hold the record for the highest critical temperature among superconductors at ambient pressure and are thus termed \textit{high-temperature superconductors} compared to simple alloys. Working with cuprates offers the advantage of utilizing liquid nitrogen for cooling instead of liquid helium. Additionally, cuprates possess a more intricate chemical composition than metallic alloys, leading to more complex phenomena. Another family often regarded as "high temperature" comprises iron-based materials, represented by the stars. This family initially surprised the scientific community, as iron, a typical magnetic material, was found to be a crucial ingredient for superconductivity. Due to their intrinsic multi-orbital structure, the theoretical framework used for simple alloys needed expansion to accommodate additional degrees of freedom and alternative electron-based pairing mechanisms.

Intriguing and exotic families of superconductors include ferromagnetic and chiral varieties. Ferromagnetic superconductors refer to materials where magnetism and superconductivity coexist, albeit with the magnetism originating independently of superconductivity. In contrast, chiral superconductors exhibit magnetic properties that emerge from the superconducting state itself. Additionally, families such as heavy-fermions, Heusler's compounds, and 2D materials each possess unique properties that are subjects of intense study. Fig. \ref{fig:history_of_sc} is not exhaustive; it serves merely as a sample of the history of superconductivity.

Where might we discover superconductors in the future? The vision of superconducting power grids and levitating vehicles remains a tantalizing prospect, contingent upon the speculative discovery of a cost-effective ambient-temperature superconductor. In the interim, other less speculative applications include high magnetic field energy storage systems, fusion reactors, and quantum sensors. Among the most promising and serendipitous applications of superconductivity is quantum computing. The physical realization of a qubit network, essential for quantum computers, appears most achievable through superconducting devices, offering robust coherence maintenance for qubits. Additionally, there is a realistic prospect of developing superconducting electronics. Rather than relying on semiconductor chips, all common operations could be executed on superconducting chips. To actualize this, superconducting equivalents of transistors, diodes, capacitors, and inductors are necessary, a field currently under active research.

However, in a philosophical standoff with many fundamental research investors, the real impact of fundamental science can hardly be predicted. New solutions create new interesting problems, which historically have their long-term value systematically underestimated.

\subsection{What? \label{sec:what2}}

Textbooks usually introduce superconductors through their properties \cite{ashcroft76,Kittel2004}, without providing a clear definition. 
The emphasis underscores the empirical properties, which include:
\begin{enumerate}
\item The absence of measurable resistivity without a magnetic field;
\item Exhibiting perfect diamagnetism in the presence of weak magnetic fields;
\item Manifesting an energy gap within the material's energy spectrum.
\end{enumerate}
These properties, together with the fact the superconducting state begins at a thermodynamic phase transition at a certain critical temperature $T_c$, are typically heralded as benchmarks for identifying new superconductors. Demonstrating these characteristics is often seen as a critical step in substantiating the discovery of novel superconducting materials. 
However, while these attributes are emblematic of conventional superconductors, they do not universally apply, thus lacking in both explanatory and predictive capacities. Consequently, this approach to defining superconductors falls short of offering a good definition.

For instance, some superconductors host topological defects, such as vortices or skyrmions, which may be either extrinsic or intrinsic. These topological objects exhibit peculiar relaxation dynamics that typically lead to residual resistivity. Some superconductors may lack diamagnetic behaviour and instead appear paramagnetic. The presence of an energy gap can be obscured by factors such as magnetic impurities or a lack of perfect macroscopic coherence. A material can be superconducting even if it does not exhibit any of the empirical properties mentioned above. These properties are merely consequences of the deeper physical phenomenon that defines superconductivity. Admittedly, the one-line definition of superconductivity is likely incomprehensible to most undergraduate physics students, which is why introducing superconductors through their characteristic phenomena seems more didactic.

Based on Refs. \cite{Annett2004,Altland2023}, we now attempt a one-sentence definition. Don't worry about its technicality. We will unpack its meaning shortly. 
\begin{align}
& \text{\textit{Superconductivity is a macroscopic coherent quantum state of pairs of fermions that breaks
}}  \notag \\
& \text{\textit{the $U(1)$ phase invariance symmetry.}} \label{eq:definition}
\end{align}
We now explain this definition in three steps: the quantum state, the $U(1)$ symmetry, and the coherent state.

\section{The order parameter\label{sec:order_parameter}}

First, superconductivity is a quantum state. 
This means that superconducting phenomena cannot be explained by a classical theory.
Several topics in solid state theory have a useful classical description, such as the Drude model of resistivity. This is not the case for superconductivity. 
The state is described by a complex wavefunction $\Psi(\boldsymbol{r},t)$, which has an amplitude $\Psi_0(\boldsymbol{r},t)=|\Psi(\boldsymbol{r},t)|$ and a phase $\varphi(\boldsymbol{r},t)$, such that
\begin{align}
\Psi(\boldsymbol{r},t)=\Psi_0(\boldsymbol{r},t)e^{i\varphi(\boldsymbol{r},t)}.
\label{eq:wavefunction}
\end{align}
Eq. \eqref{eq:wavefunction} is a general wavefunction that applies not only to superconductors. In a superconductor, however, the phase and amplitude are rigid. They might fluctuate, but if they do, the oscillations are usually small around an equilibrium value.

Second, the quantum state is macroscopic, which means that the field $\Psi(\boldsymbol{r},t)$ describes the entire superconductor.
The simplest microscopic quantum theory of superconductivity was obtained by Bardeen, Cooper and Schrieffer, which is simply known as \textit{BCS theory} \cite{bcs1957}. There, they considered the simplest case where $\Psi(\boldsymbol{r},t)$ has no spatial or time variations, such that one may write $\Psi=\Psi_0 e^{i\varphi}$. Then, $\Psi$ is a fixed complex number, also referred to as the superconducting \textit{order parameter}. 
For simplicity, we henceforth ignore the spatial and temporal variations of the order parameter. However, it is important to note that our overarching definition of superconductivity is applicable even in the more general cases where such variations may be considered.

So far, we have understood that the superconducting state can be represented by a fixed complex number that we may view as an \textit{arrow} in the complex plane. 
The arrow has a fixed amplitude $\Psi_0$ and a fixed phase $\varphi$. The angular part $e^{i\varphi}$ is a unit vector over the unit circle. The unit circle is parametrized by a single parameter: the angle $\varphi$. For this reason, one names the symmetry group of the unit circle as $U(1)$. Therefore, saying that a system has $U(1)$ symmetry is just a fancy way of saying that it has the same symmetry as a circle. Does the arrow $\Psi\sim e^{i\varphi}$ have the symmetry of a circle? We can check. If we rotate a circle it remains unchanged. However, if we rotate $e^{i\varphi}$ by an arbitrary angle in the complex plane, it generally changes its direction. Therefore, a rigid $\Psi$ \textit{is not} $U(1)$ symmetric. We then say that $\Psi\neq 0$ breaks the $U(1)$ phase invariance symmetry.
A nuanced yet crucial aspect of quantum mechanics is that the specific value of $\varphi$ is not directly observable; only relative phases can be measured. Despite this lack of experimental access to $\varphi$, it is its fixed nature that leads to the hallmark properties of superconductors mentioned at the beginning of Sec. \ref{sec:what2}.

Third, the superconducting wavefunction represents a coherent state of fermion pairs, called \textit{Cooper pairs}. 
In condensed matter physics, these fermions are electrons, though superconductivity can emerge in astrophysical objects like neutron stars. 
In a superconductor, there is always an effective interaction that serves as a mechanism that pairs up electrons. 
These paired-up electrons can be thought of as being in a bound state, the collection of which is described by $\Psi$, which has the same phase $\phi$ for the entire collection. For this reason, we call it a coherent state of pairs of electrons.

To comprehend the formation of this global wave function, it's essential to recognize that the notion of tightly bound, localized pairs of electrons is merely a simplified depiction. To clear a common misconception we appeal to Bardeen \cite{bardeen1973}:
\begin{quote}
    \textit{
The key thing is pairing, not pairs. There is no localized pairing of electrons into "pseudo-molecules" which obey Bose statistics. Although this analogy is often used, particularly by Bogoliubov and coworkers, I think it is misleading. The reason for the condensation is not Bose-Einstein statistics, but it comes from the exclusion principle; pairing allows one to make best use of the available phase space to form a coherent low-energy ground state. 
    }
\end{quote}
Typically, the average volume occupied by a Cooper pair exceeds that of a single electron by a considerable margin \cite{Tinkham2004,Ketterson1999}, resulting in pair overlap.
With this picture, it is more appropriate to think of the superconducting state as many electrons coming in and out of these bound pairs, always switching partners in a complex dance.
For temperatures above the $T_c$, the electrons are "uncoordinated", and no superconducting state is achieved, that is, no coherent state of fermion pairs is achieved. 
On the other hand, if the conditions are right (if the temperature is low enough, no exceedingly high external magnetic field, etc.) this dance of multiple electrons exchanging partners creates a macroscopic “cooperation” of the quantum state that acquires a unified phase and amplitude, just as a group of dancers will move as one when performing a well-trained choreography.

The nature of the interaction leading to Cooper pairing can be diverse. The most famous example is the electron-phonon interaction. In this case, the excitations of the ionic lattice vibrations sway the electrons just in the right way to move in tandem. Because of the regularity of the crystal which is communicated through local interactions, Cooper pairing occurs for the entire condensate in a synchronized way. Other possible causes of a superconducting interaction may come from purely electronic mechanisms, such as spin fluctuations. In real materials, many possible pairing mechanisms coexist, and identifying the dominant interaction is not always an easy task.

That said, we see that the fundamental unit of superconductivity is a pair of electrons. For those with a background in quantum mechanics, it's known that pairs of interacting spin-$1/2$ particles form either a singlet or a triplet state. This extends to the macroscopic quantum state of a superconductor, which can manifest as a singlet, a triplet, or, in certain exotic situations, a superposition of both singlet and triplet states. To simplify our discussion, we will not delve into the specifics of spin configurations.

Undergraduate projects in the field of superconductivity frequently encounter a significant learning curve. This challenge primarily stems from the complexity of quantum theory for many-particle systems, which is more aptly conveyed through the use of field operators—a subject typically reserved for graduate-level study. Consequently, projects that involve BCS theory are often restricted to students in their final undergraduate year.
Nonetheless, substantial progress can be achieved by adopting the definition presented in \eqref{eq:definition}, while temporarily setting aside the intricate microscopic underpinnings of this definition. 
By acknowledging that superconductivity can be characterized by a rigid complex order parameter, students can employ undergraduate-level thermodynamics concepts to phenomenologically analyze the system's free energy. This approach effectively circumvents the daunting learning curve, allowing the deeper theoretical exploration to be deferred until graduate studies.

\section{Symmetries of unconventional superconductors \label{sec:unconventional}}

Symmetry is the property of a system that remains unchanged under certain transformations. In physics, the role of symmetry is to provide us with selection rules, that is, a specification of what is possible and impossible. This enables us to discard many microscopic theories based on symmetry alone and helps us to pinpoint the class of microscopic theories that remain good candidates. 
For some practical purposes, the microscopic theories may even be unnecessary. 
Here we
discuss three transformations: phase changes, spatial rotations, and time-reversal symmetry.

\subsection{Broken gauge symmetry}

In the case of the $U(1)$ symmetry, the transformations are on the phase of the wavefunction, which is why it is also called a gauge symmetry.  
The terms \textit{gauge symmetry} and \textit{phase invariance symmetry} are used synonymously.
In Sec. \ref{sec:what} we already mentioned that one of the prime consequences of gauge symmetry breaking is the phenomena of superconductivity, which occurs at a thermodynamic phase transition, and accompanies other properties such as the expulsion of the magnetic induction from the interior of the material under the action of weak magnetic fields. This phenomenon is also known as the \textit{Meissner effect}. According to the definition \eqref{eq:definition}, a superconductor necessarily lacks gauge symmetry, since the phase acquires rigidity. The strength of symmetry breaking is quantified by the amplitude $\Psi_0$, because $\Psi_0^2$ is proportional to the density of electrons that condense into the coherent state. 
The only way to restore gauge symmetry is with $\Psi=0$, for which the material is in its normal (non-superconducting) state.

\subsection{Spatial rotations}

In a more general context, the phase $\varphi(\boldsymbol{r},t)$ can exhibit spatial modulation. As long as $\varphi(\boldsymbol{r},t)$ remains rigid over time, the system qualifies as a superconductor because the $U(1)$ symmetry is still broken. 

Let us first consider an example where no spatial symmetry is broken. 
Spatial rotations refer to transformations that change the orientation of objects or coordinate systems in space without altering their shape or size.
Suppose $\Psi(\boldsymbol{r})=s$, where $s$ represents a complex number independent of $\boldsymbol{r}$. Given that $s$ is merely a complex number, it is invariant under spatial rotations. Drawing a parallel to atomic physics, one refers to $\Psi(\boldsymbol{r})=s$ as an $s$-wave order parameter, reflecting the symmetry characteristics of an $s$-orbital. Spatial rotations leave the $s$-wave unchanged, indicating that $s$-wave order parameters break the $U(1)$ symmetry without violating any additional symmetries. Superconductors that solely break the $U(1)$ symmetry, without affecting other symmetries, are often termed \textit{conventional superconductors}.

In contrast, an \textit{unconventional superconductor} is characterized by an order parameter that breaks additional symmetries beyond $U(1)$. The past three decades have witnessed significant advancements in the field of unconventional superconductivity, leading to a somewhat paradoxical situation. The term \textit{unconventional} might imply rarity or deviation from the norm according to its standard definition. However, unconventional superconductors are becoming increasingly prevalent, not rarer. In this context, \textit{unconventional} is better understood as a reduction in symmetry. We may generalize the definition:
\begin{align}
& \text{\textit{Unconventional superconductivity is a macroscopic coherent quantum state of pairs of fermions that
}}  \notag \\
& \text{\textit{breaks additional symmetries beyond $U(1)$.}}
\label{eq:def_unc}
\end{align}

As a first symmetry-reduced example, consider a wavefunction that has the same spatial symmetries as an atomic $p$-orbital, $\Psi(\boldsymbol{r})=p(\boldsymbol{r})$. We could write down the specific mathematical expression for the spherical harmonic for $p(\boldsymbol{r})$, or any other function that has the symmetries of a $p$-orbital. However, for our purposes, we find it unnecessary to get distracted with mathematical details.  The important thing is to remember that if we rotate a $p_x$-orbital around the $z$ axis by $\pi$, then the transformed configuration has the opposite sign as compared to the initial configuration. For this reason, an order parameter that shares the symmetries of a $p$-orbital is frequently called a $p$-wave. Since it is not rotationally symmetric, a $p$-wave is an unconventional order parameter. A similar analysis would apply to $d$- and $f$-waves. 

The orbital physics analogy is used extensively in specialized literature. However, we stress that the analogy only serves as a proxy for the actual symmetry-reduced functions. Orbitals, or spherical harmonics, arise by studying for instance the hydrogen atom, which has full rotational symmetry. Then, angular momentum $l$ is a good quantum number, such one establishes the nomenclature $s$ ($l=0$), $p$ ($l=1$), $d$ ($l=2$), and so forth. 
The analogy fails because a crystalline solid is not fully rotationally symmetric. Spatial symmetries are discrete, and angular momentum is not a good quantum number in solids. Yet, the proxy orbital analogy renders itself useful in a simplified discussion.

\subsection{Time-reversal symmetry}\label{subsec:TRS}

Time-reversal symmetry refers to a property of a system where it remains invariant when the direction of time is reversed. 
This concept might seem abstract since, unlike spatial rotations, reversing time is not a physical action we can perform. 
For illustration, take the equation of motion $x(t) = x_0 + vt$. If time is reversed, the trajectory becomes $x(-t) = x_0 - vt$, mirroring the original path but in the opposite direction. Although directly reversing time, $t \rightarrow -t$, is not possible, we can obtain the same equation of motion by inverting velocity, $v \rightarrow -v$. This approach allows us to investigate time-reversal symmetry by observing whether reversing motion (i.e., changing $v \rightarrow -v$) retraces the original trajectory backwards. If the path diverges from the original one upon this inversion, it indicates a violation of time-reversal symmetry.

\begin{figure}
\includegraphics[width=0.35\textwidth]{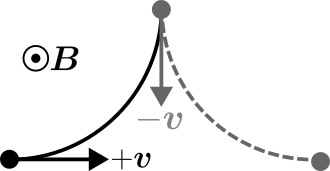}
\centering
\caption{
Reversal of motion of an electron (disks) under the influence of a magnetic field $\boldsymbol{B}$ experiencing the Lorentz force $\boldsymbol{F}\sim \mp\boldsymbol{v}\times \boldsymbol{B}$. The dashed-gray arc illustrates the reversed trajectory. 
\label{fig:trsb}}
\end{figure}

The paradigmatic example of time-reversal symmetry breaking is the motion of an electron under the influence of a magnetic field; see Fig. \ref{fig:trsb}.
In the illustration, the electron (black disk) is embedded in a uniform magnetic induction field $\boldsymbol{B}$ and hence experiences a Lorenz force $\boldsymbol{F}\sim -\boldsymbol{v}\times \boldsymbol{B}$. After moving a quarter of a circumference, we reverse the electrons' velocity $\boldsymbol{v}\rightarrow -\boldsymbol{v}$. This also changes the sign of the Lorenz force, which causes the reversed (dashed-gray) trajectory to be different. 
In condensed matter physics, the presence of time-reversal symmetry breaking usually implies that there is a magnetic field, magnetization, or another effective field that works as if it were a type of magnetic field. 

How does time-reversal symmetry affect wavefunctions? 
Wavefunctions obey the Schrödinger equation $i\hbar\,\partial_t\Psi(\boldsymbol{r},t) =\mathcal{H}\Psi(\boldsymbol{r},t)$. If the Hamiltonian $\mathcal{H}$ does not explicitly depend on time, then the solution has the form \cite{Griffiths2018}
\begin{align}
\Psi(\boldsymbol{r},t)=e^{-\frac{i}{\hbar}\mathcal{H}t}\Psi(\boldsymbol{r},0).
\end{align}
Note that the time-reversed wavefunction $\Psi(\boldsymbol{r},-t)=\Psi^*(\boldsymbol{r},t)$. Therefore, in quantum mechanics (ignoring spin), the time-reversed wavefunction may simply be obtained by complex-conjugating the original one. 

Now let us apply this to superconductivity. 
A superconducting order parameters breaks time-reversal symmetry if $\Psi(\boldsymbol{r},t)\neq\Psi^*(\boldsymbol{r},t)$. It is clear that if $\Psi$ is real, then time reversal is preserved, as in the $s$-wave $\Psi=s$.  
To break time reversal, the order parameter should have both a real part and a relative imaginary part, such as $p+ip$ or $s+if$. 
Materials with relative complex components are called \textit{chiral superconductors}
\cite{Ramires2022}.

We must alert the reader that there are conflicting usages of the term \textit{unconventional superconductivity} in the community. 
It is rather common to hear talks or online definitions suggesting that unconventional superconductors are materials that display superconductivity which does not conform to conventional BCS theory or its extensions. This is not true, because the theory of unconventional superconductors is a rather straightforward generalization of BCS theory \cite{ueda}.
Others say that unconventional superconductors are those that have a pairing mechanism that is different from the electron-phonon interaction, and hence are not described by BCS theory. 
In our view, this is a bad definition and also reveals a common misconception. 
The misconception is that BCS theory only describes superconducting states originating from the electron-phonon interaction. However, as we mentioned before, BCS theory only specifies a generic attractive interaction, that could have any origin. There are non-phonon superconductors that are very well described by BCS theory. 
Different superconducting states become thermodynamically distinguishable not necessarily due to the pairing mechanism, but mostly because of the different symmetry properties. For this reason, we argue that the symmetry-based definition is the more consistent one.

\section{Ginzburg-Landau theory \label{sec:gl}}

The primary physical consequences of symmetry breaking include phase transitions, the rigidity of the order parameter, collective excitations, and topological defects \cite{Blundell2001}. These phenomena can be analyzed through the phenomenological Landau framework.
In statistical mechanics, understanding the Hamiltonian, $\mathcal{H}$, is crucial for calculating the partition function, $\mathcal{Z} = \mathrm{tr}\left(e^{-\beta \mathcal{H}}\right)$, where $\beta=1/(k_B T)$ is the inverse temperature. This enables the determination of the system's free energy, $\mathcal{F} = -k_B T \ln \mathcal{Z}$.
The order parameter of the system adopts a value that minimizes the free energy, which corresponds to thermodynamic equilibrium where entropy reaches its maximum. However, the challenge arises when the Hamiltonian $\mathcal{H}$ is either not fully understood or is too complex for direct analytical computation of the free energy.
Nevertheless, an understanding of the symmetries of the normal state, alongside the symmetry broken during the phase transition, can significantly constrain the possible forms of the free energy. Thus, even in the absence of detailed microscopic insights or in cases of computational complexity, the principles of symmetry can provide substantial information based on selection rules alone.

In condensed matter physics, common order parameters include density, magnetization, polarization, and the superconducting wave function. Among these, the Landau theory of ferromagnetism serves as a particularly instructive example, illustrating the core principles of phase transitions and symmetry breaking.
To introduce the fundamental aspects of Landau theory, we recall the case of ferromagnetism.

\subsection{Ferromagnetism}

In studying the transition from a high-temperature paramagnet to a low-temperature ferromagnet, we know that a magnetization $\boldsymbol{M}$ (order parameter), develops below the Curie temperature $T_c$. Based on symmetry considerations alone, we now would like to guess the form for the free energy as a function of the magnetization $\boldsymbol{M}$. 
Here, the order parameter $\boldsymbol{M}$ is a real vector, which is zero above $T_c$, and finite below. The symmetry that the magnetization $\boldsymbol{M}$ breaks is the rotational symmetry of the spins, which may be expressed in terms of the group $SO(3)$. 
Unlike for the superconducting order parameter, here, the order parameter $\boldsymbol{M}$ already corresponds to a physical observable, which is why it is perhaps more didactic than beginning with the superconducting case.

The system is mathematically defined by its Hamiltonian. Since the Hamiltonian represents the system, it must have the same symmetries as the system. The paramagnet has rotational $SO(3)$ symmetry, which implies that the Hamiltonian also must preserve $SO(3)$. Because the free energy is connected to the Hamiltonian via $\beta\mathcal{F} = -\ln \mathrm{tr}\left(e^{-\beta \mathcal{H}}\right)$, the free energy must also preserve the high-temperature symmetry. 
Supposing that the system has the symmetries of $SO(3)$, then the free energy must respect these symmetries, even below $T_c$. Below, but close to $T_c$, we may develop an expansion of the free energy in terms of the order parameter:
\begin{align}
\mathcal{F}[\boldsymbol{M}]=\mathcal{F}_0+\cancel{c_1\boldsymbol{M}}+c_2\boldsymbol{M}^2+\cancel{c_3 \boldsymbol{M}^3}+c_4 \boldsymbol{M}^4+\ldots,
\end{align}
where $\mathcal{F}_0$ is the free energy of the normal state, and the $\{c_i\}$ are real coefficients that may be derived from the Hamiltonian, or used as phenomenological coefficients. 
The first term $c_1\boldsymbol{M}$ should immediately alarm. The reason is that the free energy is a scalar, whereas $\boldsymbol{M}$ is a vector. Therefore, only even powers such as $\boldsymbol{M}^2=\boldsymbol{M}\cdot \boldsymbol{M}=|\boldsymbol{M}|^2$ are allowed in the free energy expansion. Next, we must check whether the remaining terms survive the symmetry requirements for the free energy. 
Any rotation in $SO(3)$ 
leaves $|\boldsymbol{M}|^2$ invariant, which confirms that all even powers of the order parameters are symmetry-allowed. 
Therefore, the free energy close to $T_c$ may be phenomenologically expressed as
\begin{align}
\mathcal{F}[\boldsymbol{M}]=\mathcal{F}_0+a\left(\frac{T}{T_c}-1 \right )M^2+\frac{b}{2}M^4+\ldots,
\label{eq:free_ferro}
\end{align}
where $a,b>0$ and $M=|\boldsymbol{M}|$. 
The phenomenological parameter $b$ is necessarily positive to guarantee that the free energy is bounded from below. 
From Eq. \eqref{eq:free_ferro} we see that if $T\geq T_c$, $M=0$ minimizes the free energy.
If $T<T_c$, $M>0$ minimizes the free energy. 
In the ferromagnetic state ($M>0$), spin rotation symmetry $SO(3)$, which is a symmetry of the free energy (or Hamiltonian), is broken. Obtaining the temperature dependence $M(T)$ close to $T_c$ in terms of the phenomenological parameters $(a,b)$ is now a straightforward task.

\subsection{Superconductivity}

Real materials have a crystal structure that lacks full rotational symmetry. 
As a working example, inspired by cuprates and strontium ruthenate, 
consider a system with the following discrete symmetries of a square:
\begin{itemize}
    \item $C_4(z)$: Rotations of $\pi/2$ about the $z$ axis $(x,y)\rightarrow (y,-x)$;
    \item $C_2(z)$: Rotations of $\pi$ about the $z$ axis $(x,y)\rightarrow (-x,-y)$;
    \item $C_2(x)$: Rotations of $\pi$ about the $x$ axis $(x,y)\rightarrow (x,-y)$;
    \item $C_2(d)$: Rotations of $\pi$ about the diagonal in the $x-y$ plane $(x,y)\rightarrow (y,x)$.
\end{itemize}
Then, the free energy (or Hamiltonian) of the superconductor should respect these four discrete spatial symmetries, plus $U(1)$.

In a superconductor, the order parameter is its wavefunction; see Eq. \eqref{eq:wavefunction}. For simplicity, ignoring the spatial and temporal variations, the superconducting free energy will have the form
\begin{align}
\mathcal{F}[\Psi]=\mathcal{F}_0+a\left(\frac{T}{T_c}-1 \right )\Psi_0^2+\frac{b}{2}\Psi_0^4+\ldots
\label{eq:free_superconductor}
\end{align}
The free energy cannot contain the term $c_1\Psi$, because $\Psi$ is a complex function and $\mathcal{F}$ is real. Also, only the amplitude $\Psi_0=|\Psi|$ occurs in the free energy because, in this example, the system is homogeneous. Minimization of the free energy with respect to $\Psi_0$ determines the temperature dependence $\Psi_0(T)$. Although the phase $\varphi$ of the order parameter is rigid, the minimization of the free energy cannot provide it, because only relative phases relate to observables. Since there is only one order parameter component in Eq. \eqref{eq:free_superconductor}, it is customary to choose $\varphi=0$, such that the order parameter is real. 
Since $\Psi_0$ does not depend on the spatial coordinates, any spatial symmetry operation leaves $\Psi_0$ invariant, which shows that Eq. \eqref{eq:free_superconductor} respects the symmetries of the system. 

We may readily generalize Eq. \eqref{eq:free_superconductor} for unconventional order parameters that break spatial symmetries, but there is no hope of getting an order parameter that breaks time-reversal symmetry from Eq. \eqref{eq:free_superconductor}, since this would also require a relative imaginary component. 

\section{Two-component superconductors \label{sec:two}}

Let us assume that in an
unconventional 2D superconductor, the order parameter can be represented by a two-component vector $\boldsymbol{\Psi}=(\Psi_x,\Psi_y)$. The configuration of $\boldsymbol{\Psi}$, determined by minimizing the free energy, can lead to chiral superconductivity if $\Psi_x$ and $\Psi_y$ develop a relative imaginary component, thereby breaking time-reversal symmetry. When expanding the real free energy in terms of the complex vector $\boldsymbol{\Psi}$, we must identify symmetry-permitted terms. Since $\mathcal{F}$ is a real scalar, it cannot be directly proportional to $\boldsymbol{\Psi}$. Moreover, the term $\boldsymbol{\Psi}^2=\Psi_x^2+\Psi_y^2$ is generally complex and thus also not permissible. Let's focus on identifying the first symmetry-allowed term in the free energy expansion.

\subsection{The quadratic term \texorpdfstring{$|\boldsymbol{\Psi}|^2$}{b} \label{sec:p2}}

The first good candidate is
\begin{align}
|\boldsymbol{\Psi}|^2=(\Psi_x,\Psi_y)\cdot (\Psi_x^*,\Psi_y^*)=|\Psi_x|^2+|\Psi_y|^2,
\label{eq:p2}
\end{align}
which is a real scalar, just as $\mathcal{F}$. 
We now have to check whether $|\boldsymbol{\Psi}|^2$ remains invariant under the symmetries of the free energy, in this case, $C_4(z)$, $C_2(z)$, $C_2(x)$, $C_2(d)$ and $U(1)$. 
The absolute value $|\boldsymbol{\Psi}|^2$ is certainly $U(1)$ gauge invariant, since a change of the phases of $\Psi_x$ and $\Psi_y$ does not manifest in $|\boldsymbol{\Psi}|^2$. 
The $C_4(z)$ about the $x$ axis transforms the order parameter components as $(\Psi_x,\Psi_y)\rightarrow (\Psi_y,-\Psi_x)$. Substituting these transformations in Eq. \eqref{eq:p2} shows that $|\boldsymbol{\Psi}|^2$ remains invariant under $C_4(z)$. 
The $C_2(z)$ takes $(\Psi_x,\Psi_y)\rightarrow (-\Psi_x,-\Psi_y)$,
which yields $|\boldsymbol{\Psi}|^2\rightarrow |\boldsymbol{\Psi}|^2$. 
The $C_2(x)$ rotation transforms the components as $(\Psi_x,\Psi_y)\rightarrow (\Psi_x,-\Psi_y)$, which also does not change $|\boldsymbol{\Psi}|^2$. 
Lastly, the $C_2(d)$ rotation exchanges the components $(\Psi_x,\Psi_y)\rightarrow (\Psi_y,\Psi_x)$, which also leaves $|\boldsymbol{\Psi}|^2$ intact.  
The term $|\boldsymbol{\Psi}|^2$ survived all symmetries of the system and for this reason will appear in the free energy (Eq. \eqref{eq:free_allowed}) expansion with its exclusive phenomenological parameter $(b_1)$.

\subsection{The quadratic term \texorpdfstring{$\cancel{\Psi_x\Psi_y^*\mp \Psi_x^*\Psi_y}
$}{b}\label{sec:mixed}}

Because $\boldsymbol{\Psi}$ has two components, there are other possible quadratic real combinations, namely, $\Psi_x\Psi_y^*\mp \Psi_x^*\Psi_y$. 
Following the same transformations as in Sec. \ref{sec:p2}, you may check that $\Psi_x\Psi_y^*- \Psi_x^*\Psi_y$ acquires a minus sign under $C_2(d)$ and $\Psi_x\Psi_y^* + \Psi_x^*\Psi_y$ a minus sign under $C_4(z)$. Since $\Psi_x\Psi_y^*\mp \Psi_x^*\Psi_y$ breaks at least one symmetry of the system, this quadratic combination is prohibited. We were able to state an impossibility, without detailed knowledge of the Hamiltonian.

\subsection{Quartic terms}

To enumerate all feasible fourth-order combinations, first consider the expression:
\begin{equation}
|\boldsymbol{\Psi}|^4 = |\Psi_x|^4 + |\Psi_y|^4 + 2|\Psi_x|^2|\Psi_y|^2.
\end{equation}
Drawing from the discussions in Section \ref{sec:p2}, it is evident that the term $|\boldsymbol{\Psi}|^4$ satisfies all symmetries of the square. 
Furthermore, it is important to recognize that the components $|\Psi_x|^4 + |\Psi_y|^4$ and $|\Psi_x|^2|\Psi_y|^2$ independently adhere to all symmetry requirements. This observation allows for the assignment of distinct phenomenological parameters to these components. Specifically, we opt to associate the parameter $b_3$ with $|\Psi_x|^2|\Psi_y|^2$.
While the selection of $b_3$ is not the sole possibility, as delineated in various references such as Refs. \cite{Ramires2022,annett90,Sigrist2005,VolovikGorkov1985}, any chosen parameterization leads to a model that captures the same superconducting instabilities effectively.

In the previous Sec. \ref{sec:mixed}, we saw that $\Psi_x\Psi_y^*\mp \Psi_x^*\Psi_y$ acquires a minus sign under some transformations, which prohibits the term from entering the free energy. However, the minus sign is taken care of by taking the square:
\begin{align}
\left(\Psi_x\Psi_y^*\mp \Psi_x^*\Psi_y \right )^2=\Psi_x^2{\Psi_y^*}^2+{\Psi_x^*}^2\Psi_y^2\mp 2|\Psi_x|^2|\Psi_y|^2,
\label{eq:cross}
\end{align}
which shows that this term is allowed in the free energy. 
Note that $|\Psi_x|^2|\Psi_y|^2$ is also contained in Eq. \eqref{eq:cross}, which receives its own phenomenological parameter $b_3$. 
Then we could associate an independent phenomenological parameter $b_2$ to either $\left(\Psi_x\Psi_y^*\mp \Psi_x^*\Psi_y \right )^2$ \cite{Ramires2022} or  $\Psi_x^2{\Psi_y^*}^2+{\Psi_x^*}^2\Psi_y^2$ \cite{Sigrist2005}. We choose the latter. 
We could continue the exercise for the sixth-order terms,
but if we are close enough to $T_c$, it is usually sufficient to truncate the free energy at the quartic order.

\subsection{The free energy}

We may now construct the free energy for the two-component superconductor up to the fourth order by including all the symmetry-allowed combinations with their respective phenomenological parameters:
\begin{align}
\mathcal{F}[\boldsymbol{\Psi}]=\mathcal{F}_0(T)+a(T)|\boldsymbol{\Psi}|^2+\frac{b_1}{4}|\boldsymbol{\Psi}|^4
+\frac{b_2}{2}
\left(
\Psi_x^2{\Psi_y^*}^2+{\Psi_x^*}^2\Psi_y^2 
\right )
+b_3|\Psi_x|^2|\Psi_y|^2+\ldots
\label{eq:free_allowed}
\end{align}
Here $a(T)=a(T/T_c-1)$, with $a>0$.
The prefactors for the $\{b_i\}$ were chosen for later convenience. 
The free energy contains three variational parameters: the amplitudes $|\Psi_x|$ and $|\Psi_y|$, and the relative phase $\alpha=\varphi_y-\varphi_x$. 
To see this explicitly, we may write the components as $\Psi_x=|\Psi_x|e^{i\varphi_x}$ and $\Psi_y=|\Psi_y|e^{i\varphi_y} $, such that Eq. \eqref{eq:free_allowed} updates to 
\begin{align}
\mathcal{F}[|\Psi_i|,\alpha]=
\mathcal{F}_0(T)+a(T)
\left(|\Psi_x|^2+|\Psi_y|^2 \right )
+\frac{b_1}{4}\left(|\Psi_x|^2+|\Psi_y|^2 \right )^2
+\left(b_2\cos(2\alpha)+b_3 \right )|\Psi_x|^2|\Psi_y|^2+\ldots,
\label{eq:to_minimize}
\end{align}
where the variational parameters $(|\Psi_x|,|\Psi_y|,\alpha)$ are now explicit.
To find the equilibrium values for the variational parameters $\boldsymbol{\Psi}_\mathrm{min}=(|\Psi'_x|,|\Psi'_y|,\alpha_\mp)$, we minimize the free energy in Eq. \eqref{eq:to_minimize} with respect to $\boldsymbol{\Psi}_\mathrm{min}$ to obtain 
\begin{align}
|\Psi'_x|^2=|\Psi'_y|^2=\frac{a\left(1-\frac{T}{T_c} \right )}{b_1\mp b_2+b_3},\quad \alpha_\mp=
\begin{cases} 
\frac{\pi}{2} \\
\pi
\end{cases},\quad \Rightarrow\quad
\mathcal{F}[\boldsymbol{\Psi}_\mathrm{min}]=\mathcal{F}_0(T)-\frac{a^2\left(1-\frac{T}{T_c} \right )^2}{b_1\mp b_2+b_3}.
\label{eq:solutions}
\end{align}
From the solutions for the amplitudes in Eq. \eqref{eq:solutions}, we recognize that we must require $b_1\mp b_2+b_3>0$ simultaneously for the phenomenological parameters, which guarantees that the free energy is bounded from below. 
If the free energy were not bounded from below, this would equivalently mean that there is no maximum bound for entropy, which would imply that that system would not equilibrate. 
The two inequalities $b_1\mp b_2+b_3>0$ give rise to two distinct superconducting phases where the amplitudes of the components are equal, but with a phase difference of $\alpha_-=\pi/2$ or $\alpha_+=\pi$. 
Feeding these phases into $\boldsymbol{\Psi}_\mathrm{min}=(\Psi_x',\Psi_y')$ gives us a two-component chiral phase $(1,i)$ and a two-component real phase $(1,-1)$, respectively. 
Substituting the solution $\boldsymbol{\Psi}_\mathrm{min}$ back into the free energy gives us the minimal free energy value $\mathcal{F}[\boldsymbol{\Psi}_\mathrm{min}]$ in Eq. \eqref{eq:solutions}, which, for $a>0$ and $b_1\mp b_2+b_3>0$ gives $\mathcal{F}[\boldsymbol{\Psi}_\mathrm{min}]<\mathcal{F}_0(T)$.

\begin{figure}
\includegraphics[width=0.8\textwidth]{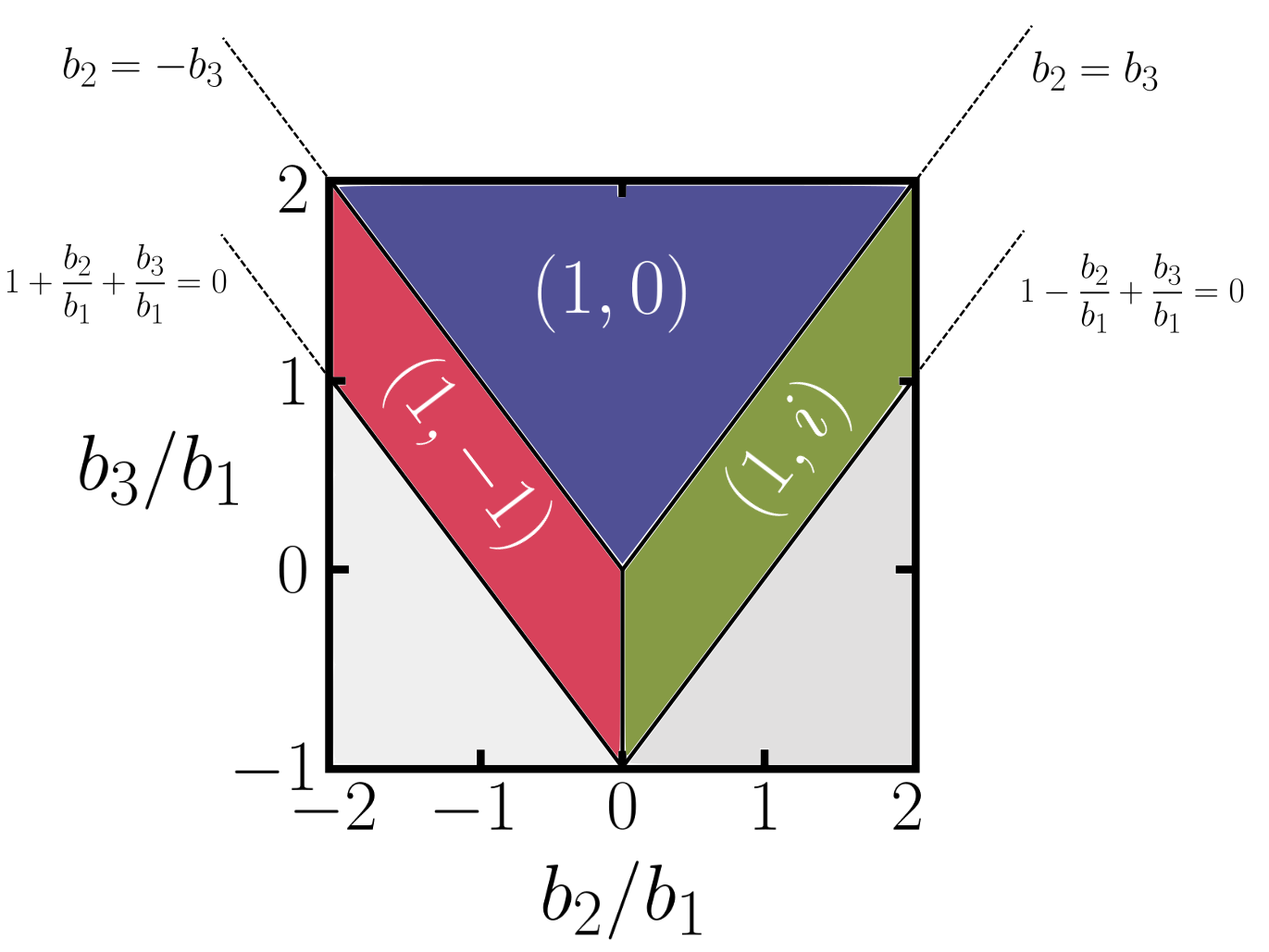}
\centering
\caption{
Superconducting phases of the free energy in Eq. \eqref{eq:free_allowed}. The blue region $(1,0)$ stabilizes a single component order parameter. 
The red region $(1,-1)$ is a real two-component order parameter with relative phase $\alpha_+=\pi$.
The green region $(1,i)$ is a chiral two-component order parameter with relative phase $\alpha_-=\pi/2$. 
The gray region is unstable, since in this region, the energy is unbounded from below ($b_1\mp b_2+b_2<0$). 
\label{fig:diagram}
}
\end{figure}

Do the phases $(1,-1)$ and $(1,i)$ exhaust all the possibilities? No. 
We must also entertain the possibility that one-component superconductivity is more favourable, which, due to the symmetry of the free energy in Eq. \eqref{eq:to_minimize} is not contemplated in the solutions in Eq. \eqref{eq:solutions}. Therefore, let us examine the value that the free energy acquires if one of the components vanishes, say $\Psi_y=0$. Then, we find that the minimum value for the single-component free energy is 
\begin{align}
    \mathcal{F}_{(1,0)}=\mathcal{F}_0(T)-\frac{a^2}{b_1}\left(1-\frac{T}{T_c} \right )^2.
    \label{eq:10}
\end{align}
If $\mathcal{F}_{(1,0)}<\mathcal{F}[\boldsymbol{\Psi}_\mathrm{min}]$, this means that single-component superconductivity is more favourable than a two-component phase. The inequality $\mathcal{F}_{(1,0)}<\mathcal{F}[\boldsymbol{\Psi}_\mathrm{min}]$ leads to the additional condition $\mp b_2+b_3>0$, which, when satisfied, stabilizes the one-component phase $(1,0)$; see dashed lines in Fig. \ref{fig:diagram}.

The same analysis could have been performed using various other parametrizations. Another popular parametrization has a Bloch-sphere form $\boldsymbol{\Psi}=(\Psi_x,\Psi_y)=\Psi_0\left(\cos\theta,e^{i\varphi}\sin\theta  \right )$, such that the three variational parameters are now $\{\Psi_0,\theta,\varphi\}$. This would lead to another equivalent free energy given by
\begin{align}
\mathcal{F}[\Psi_0,\theta,\varphi]=\mathcal{F}_0(T)+a(T)\Psi_0^2
+\frac{1}{4}\left[b_1+\sin^2(2\theta)\left(b_3+b_2\cos(2\varphi) \right ) \right ]\Psi_0^4+\ldots
\label{eq:alternative}
\end{align}
Using Eq. \eqref{eq:to_minimize} or Eq. \eqref{eq:alternative} is a matter of taste. 
Here we mention the advantage and disadvantage of using Eq. \eqref{eq:alternative} relative to Eq. \eqref{eq:to_minimize}. 
The main advantage of the $\{\Psi_0,\theta,\varphi\}$ parametrization is that the single-component phase $(1,0)$ is easy to spot, 
which occurs for $\sin^2(2\theta)=0$, for which $\theta = n\pi/2,\, n\in \mathbb{Z}$. 
The main disadvantage is that the analytical solution for the minimized parameter $\Psi_0'$ is more complicated than Eq. \eqref{eq:solutions}.

\subsection{Phase diagram}

The free energy model in Eq. \eqref{eq:free_allowed} gives three distinct superconducting phases. 
A one-component phase $(1,0)$, a real two-component phase $(1,-1)$, and a chiral phase $(1,i)$. 
We now wish to map out a phase diagram for these phases in the parameter space spanned by $b_3/b_1\times b_2/b_1$. 
For this, for each pair in the parameter space, we must compare the free energies, and check which phase has the lowest free energy value. The single-component phase has the free energy given by Eq. \eqref{eq:10}. From Eq. \eqref{eq:solutions}, we see that the two-component phases have free energies given by
\begin{align}
\mathcal{F}_{(1,-1)}=\mathcal{F}_0(T)-\frac{a^2\left(1-\frac{T}{T_c} \right )^2}{b_1+ b_2+b_3},\quad 
\mathcal{F}_{(1,i)}=\mathcal{F}_0(T)-\frac{a^2\left(1-\frac{T}{T_c} \right )^2}{b_1 - b_2+b_3}.
\end{align}
Therefore, by comparing the three free energies $\mathcal{F}_{(1,0)}$, $\mathcal{F}_{(1,-1)}$ and $\mathcal{F}_{(1,i)}$, we can determine the stable phase, which we plot in Fig. \ref{fig:diagram}.

According to our definition of an unconventional superconductor in \eqref{eq:def_unc}, all of the superconducting phases in Fig. \ref{fig:diagram} are unconventional. 
The single-component phase $(1,0)$ is not equivalent to a conventional superconductor described by the free energy in Eq. \eqref{eq:free_superconductor}. 
To see this consider for instance the $C_4(z)$ rotation that takes 
$(\Psi_x,\Psi_y)\rightarrow(\Psi_y,-\Psi_x)$. This transforms $(1,0)\rightarrow (0,-1)$, which shows that even though the order parameter has a single component, it does not have the same spatial symmetries as the free energy. This makes the $(1,0)$ phase unconventional because it breaks spatial symmetries in addition to $U(1)$. 
The $(1,-1)$ also breaks $C_4(z)$ because $(1,-1)\rightarrow (-1,-1)$. This changes the relative phases between the components, which thus corresponds to a different superconducting state. 
However, for the chiral phase $(1,i)$, the $C_4(z)$ rotation transforms $(1,i)\rightarrow (i,-1)$.
We have the freedom to shift the phase by $-i(i,-1)=(1,i)$, which retains the relative phase! The chiral phase retains $C_4(z)$ symmetry. We may also check that the chiral phase respects all other spatial symmetries ($C_4(z)$, $C_2(z)$, $C_2(x)$, $C_2(d)$) of the free energy; see Tab. \ref{tab:symmetry}. The chiral order parameter retains the spatial symmetries, but it breaks time-reversal symmetry, which also renders it unconventional according to the definition. 

\begin{table}
\centering
\begin{tabular}{@{}lllllllll@{}}
\toprule
Phase  &                   &  & $C_4(z)$                         & $C_2(z)$                         & $C_2(x)$                       & $C_2(d)$                        &  & $\mathcal{T}$                   \\ \midrule
       & $(\Psi_x,\Psi_y)$ &  & $(\Psi_y,-\Psi_x)$               & $(-\Psi_x,-\Psi_y)$              & $(\Psi_x,-\Psi_y)$             & $(\Psi_y,\Psi_x)$               &  & $(\Psi_x^*,\Psi_y^*)$           \\
       &                   &  &                                  &                                  &                                &                                 &  &                                 \\
Single & $(1,0)$           &  & {\color{green} $(0,-1)$}  & {\color{blue} $(-1,0)$}  & {\color{blue} $(1,0)$} & {\color{green} $(0,1)$}  &  & {\color{blue} $(1,0)$}  \\
Two    & $(1,-1)$          &  & {\color{green} $(-1,-1)$} & {\color{blue} $(-1,1)$}  & {\color{green} $(1,1)$} & {\color{blue} $(-1,1)$} &  & {\color{blue} $(1,-1)$} \\
Chiral & $(1,i)$           &  & {\color{blue} $(i,-1)$}  & {\color{blue} $(-1,-i)$} & { $(1,i)$} & { $(1,i)$}  &  & $(1,-i)$                        \\ \bottomrule
\end{tabular}
\caption{
Transformation properties of the order parameters under spatial transformations and time-reversal $\mathcal{T}$. The blue entries are equivalent to the original order parameter by a phase shift. The green entries indicate a broken symmetry.}
\label{tab:symmetry}
\end{table}

\section{Discussion \label{sec:discussion}}

To our knowledge, chiral or magnetic superconductors were first conjectured by Volovik and Gor'kov in 1984 \cite{VolovikGorkov1985}, initially perceived as too exotic to naturally occur in materials. However, experimental physics has since made significant strides, with the discovery of several strong candidate materials exhibiting superconducting states breaking time-reversal symmetry. For comprehensive introductions and reviews on the subject, we recommend consulting references such as \cite{Ramires2022,Ghosh2020,Wysokiński2019}.

In superconductors, the breaking of time-reversal symmetry can occur either intrinsically or extrinsically. In the intrinsic case, time-reversal symmetry is spontaneously broken by the superconducting state itself, without any external intervention. One possibility as to how this can be manifested, is as rotating surface currents that develop spontaneously due to the imaginary component of the order parameter, making the material display distinct magnetic properties. On the other hand, time-reversal symmetry is extrinsically broken when a magnetic field is applied to a material, causing a transformation of an initially non-chiral order parameter into a chiral one. For example, monolayer NbSe$_2$ is believed to exhibit single-component $s$-wave superconductivity \cite{mockli2019}. However, due to the absence of inversion symmetry in its crystal structure, which is a crucial spatial symmetry, the system theoretically allows for the coexistence of singlet and triplet order parameters. Although there is no triplet order parameter in the absence of a magnetic field, the application of a magnetic field induces an $if$ triplet component, resulting in the emergence of a chiral $s+if$ state in the order parameter. Ongoing experiments are actively investigating this chiral phase \cite{Quay2022}.
One paradigmatic example of intrinsic time-reversal symmetry breaking is Sr$_2$RuO$_4$ \cite{maeno2024}, an unconventional superconductor with a complex history of theoretical and experimental investigations. Despite ongoing debates regarding the precise form of its order parameter, the scientific community appears to be converging towards the consensus that Sr$_2$RuO$_4$ hosts a chiral singlet order parameter.

\subsection{Where to begin?}

For those interested in exploring more about superconductivity, we recommend a selection of modern resources. Beginners may find the books by Annett \cite{annett90} and Tinkham \cite{Tinkham2004} to be excellent starting points. For a more encyclopedic introduction, the book by Ketterson and Song \cite{Ketterson1999} is recommended, which draws inspiration from the seminal work of de Gennes \cite{DeGennes2018}.
Intermediate readers may benefit from the works of Kita \cite{Kita2015}, Tsuneto \cite{Tsuneto1998}, and Combescot \cite{Combescot2022}.

Accessing material on unconventional superconductivity can be particularly challenging. A foundational text in this area is the book by Mineev and Samokhin \cite{Mineev1999}, along with Kita \cite{Kita2015}. Due to the scarcity of Ref. \cite{Mineev1999}, the school papers by Sigrist \cite{Sigrist2005,Sigrist2009} serve as excellent alternatives. Another seminal resource in the field is the review paper in Ref. \cite{ueda}.

Superconductivity research utilizes various methods. The mean-field Bogoliubov-de Gennes method \cite{Ketterson1999,Zhu2016,DeGennes2018} is an accessible yet powerful approach for simulating superconductors in real space, despite its approximate nature. The quantum Monte Carlo technique offers an exact alternative but is limited by the sign problem for certain Hamiltonians, its time-consuming nature, and its inability to simulate large systems. The quasi-classical approximation, detailed in \cite{Kita2015,Kopnin2001}, is effective for studying real-space problems both numerically and analytically, provided the order parameter's energy scale is significantly smaller than the Fermi energy. 
An introduction to path integral methods applied to superconductivity may be found in Ref. \cite{Tempere2012}.
For an introduction to field theoretical methods, we suggest Refs. \cite{Altland2023,Coleman2015,Bruus2004}.

\section*{Acknowledgements}

Both authors thank the support from CNPq.

\end{document}